\documentclass[aps,pra,twocolumn,superscriptaddress,runinaddress,floatfix,preprintnumbers,showpacs,nofootinbib,preprintnumbers]{revtex4-1}

\usepackage{dcolumn}
\usepackage{amssymb}
\usepackage{amsmath}
\usepackage{amsthm}
\usepackage{latexsym}
\usepackage{graphicx}
\usepackage{color}
\usepackage{ulem}

\newcommand{\ket}[1]{\vert #1 \rangle}
\newcommand{\bra}[1]{\langle #1 \vert}
\newcommand{\hil}{\mathcal{H}}

\begin{document}


\preprint{ADP-11-45/T767}

\title{Limits on the Observable Dynamics of Mixed States}
 

\author{Cael L. Hasse}
\email{Electronic address: cael.hasse@adelaide.edu.au}
\affiliation{Special Research Centre for the Subatomic Structure of Matter and Department of Physics, University of Adelaide 5005, Australia.}

\begin{abstract}
It is shown that the observability of a large class of operations on mixed states is fundamentally limited.  We consider trace preserving, unital operations.  This class includes unitary and perfect premeasurement operations.  An upper bound on the trace distance between an untransformed state and a state transformed by one of these operations is derived. The bound is dependent only on the purity of the state.  In the case of maximal mixedness, the bound implies all operations of this class are unobservable.
\end{abstract}

\maketitle

\section{Introduction}

Given a particular quantum state that is subjected to a class of operations on the state, does the state change? And if so, how observable is that change? This question is related to the study of the preservation of information subject to quantum processes \cite{Chuang:1997, Nielsen:1996, Gilchrist:2005} by Nielsen {\it et al}. Their work focuses on the slightly different question of quantifying changes to quantum states given specific operations. By asking our initial question instead of that raised by Nielsen {\it et al.} the properties of the state are emphasized. We shall in particular explore the relationship between structural properties of a state (i.e., entanglement) and its dynamics.

There is also a foundational motivation for this. Since the inception of quantum mechanics (QM), there has been an uneasy dichotomy between two points of view: is QM a fundamental description of nature or merely an algorithm to calculate probabilities for outcomes of experiments? The friction between these two viewpoints comes from the manifestly non-classical phenomena QM predicts, constrains or allows. This list of phenomena includes interference, the uncertainty principle, non-locality through entanglement (Bell inequalities) or otherwise \cite{Bennett:1999}, quantum teleportation and no-cloning theorems \cite{Wootters:1982}. The relationships between these phenomena remain unclear \cite{Barrett:2007ip, Oppenheim:2010, Popescu:1994, Gross:2010mc}. Our interest lies in how the structure of the allowed states of a quantum theory constrains its dynamics \cite{Barrett:2007ip, Gross:2010mc}. We suggest that the results of this paper may be extrapolated to probabilistic theories
  more general than quantum mechanics.

We utilize a symmetry of quantum states called envariance \cite{Zurek:2005pr} which emerges dynamically due to their entanglement structure. This symmetry is a consequence of the tensor product structure (TPS) of quantum states%
\footnote{It has been shown \cite{Barrett:2007ip} that TPSs are a generic feature of probabilistic theories with subsystems. Some form of envariance may then exist in such theories as well.}.
In classical deterministic theories, Cartesian products are used to define assemblies of subsystems rather than tensor products and so a symmetry equivalent to envariance does not exist. 

We shall use an operational definition of the observability of the change of a
quantum state: the trace distance (defined in Sec.~V) between the transformed
and untransformed state. An upper bound on this measure of the observability of the dynamics is derived. The bound is only dependent on the purity of the state. Another intriguing aspect of quantum states is the mathematical equivalence of states of subsystems of an entangled system to mixed states representing classical ensembles of quantum states. This connection allows the bound to apply to entangled systems as well. 

Envariance is defined in Sec.~II. In Sec.~III, we describe how the 
information contained in the subsystems of an entangled bipartite system can 
be less than the information contained in the whole system. It is also 
shown how the mixedness of a state constrains knowledge on all non-degenerate 
observables of that state. These two qualities of quantum states are then 
used to motivate Sec.~IV where the class of invariant operations on a 
completely mixed state is considered. This symmetry is then used in Sec.~V 
to derive an upper bound on the trace distance between an untransformed 
state and a transformed one. Concluding remarks are presented in Sec.~VI.

\section{Envariance}

Envariance \cite{Zurek:2005pr,Paris:2005ul} is a symmetry
of entangled composite systems. We define a composite system, in
general, as a state that can be decomposed in terms of eigenstates of
two or more mutually commuting observables where subsets of the total
set of mutually commuting observables completely describe
subsystems. Therefore a particle state with quantum numbers spin and
position can be considered composite, with spin and position
describing separate subsystems.

Zurek's original use of envariance was to provide a proof of Born's
rule under `very mild' assumptions. We shall be using envariance in a
different context and thus assume Born's rule from the outset.

Consider a composite system that can be decomposed into two subsystems
$\alpha$ and $\beta$, with the state $\ket{\psi} \in \hil_\alpha
\otimes \hil_\beta$. Now suppose there exist unitary operators
$U_\alpha$ and $U_\beta$, where
\begin{equation}
U_\alpha \mathrel{\mathop:}= \bar{U}_\alpha \otimes \mathbb{I}_\beta,
\label{first}
\end{equation}
with $\mathbb{I}_\beta$ being the identity on $\hil_\beta$ and
$\bar{U}_\alpha : \tilde{\hil}_\alpha \rightarrow \tilde{\hil}_\alpha$
where $\tilde{\hil}_\alpha$ is a subspace of $\hil_\alpha$, with an analogous definition for $U_\beta$. A state $\ket{\psi}$ is said to be envariant under $U_{\alpha}$ (or $U_\beta$) if the following holds:
\begin{equation}
U_\alpha U_\beta \ket{\psi} = \ket{\psi} \quad \mbox{or} \quad U_\alpha |\psi\rangle = U_\beta^\dagger|\psi\rangle.
\label{envdef}
\end{equation}
Note that from now on, we shall be considering finite dimensional
Hilbert spaces only.

Suppose we have a state of the form
\begin{equation}
\ket{\psi} = \frac{1}{\sqrt{N}}\sum^N_{j=1} e^{i\phi_j}\ket{\alpha_j}\ket{\beta_j},
\label{third}
\end{equation}
where $\left\{\ket{\alpha_j}\right\}$ and
$\left\{\ket{\beta_j}\right\}$ form orthonormal bases
for $\tilde{\hil}_\alpha$ and $\tilde{\hil}_\beta$ respectively, $N =
\textrm{dim}(\tilde{\hil}_\alpha) = \textrm{dim}(\tilde{\hil}_\beta)$
and $\phi_j$ are arbitrary phases. These states are envariant under
{\it all} unitary transformations of $\tilde{\hil}_\alpha$ (or
$\tilde{\hil}_\beta$). When $\tilde{\hil}_\alpha = \hil_\alpha$, the
state is maximally entangled (for subsystems $\alpha$ and $\beta$) and
the group of envariant transformations is the group of all unitary
transformations of $\hil_\alpha$ alone (i.e., they can be decomposed as
in Eq.~(\ref{first})).

Consider now the case of a state with Schmidt decomposition
\begin{equation}
\ket{\Omega} = \sum^n_{j=1} c_j e^{i\phi_j}\ket{\alpha_j}\ket{\beta_j},
\label{fourth}
\end{equation}
with $c_j \in \mathbb{R}^{+}$ such that $c_i \neq c_j$ for $i \neq j$, i.e., the
coefficients of the Schmidt decomposition have unequal norms. In
this case, the group of envariant transformations includes only
relative (and overall) phase changes between the components
$\ket{\alpha_i}\ket{\beta_i}$, i.e., unitaries of the form
\begin{equation}
U_\alpha := \Big(\sum^n_{j=1}e^{i\lambda_j}\ket{\alpha_j}\bra{\alpha_j} + \sum^N_{j=n+1}\ket{\alpha_j}\bra{\alpha_j}\Big)\otimes\mathbb{I}_\beta,
\end{equation}
\begin{equation}
U_\beta := \Big(\sum^n_{k=1}e^{-i\lambda_k}\ket{\beta_k}\bra{\beta_k} + \sum^M_{k=n+1}\ket{\beta_k}\bra{\beta_k}\Big)\otimes\mathbb{I}_\alpha,
\end{equation}
where $\textrm{dim}(\hil_\alpha) = N, \textrm{dim}(\hil_\beta) = M$
and $\lambda_j \in (0,2\pi)$ $\forall$ $j$. These unitaries have the desirable property
\begin{equation}
U_\alpha U_\beta \ket{\Omega} = \ket{\Omega}.
\end{equation}

The most general case is where some of the coefficients $c_i$ are
equal and others are not. For the subspaces of $\hil_\alpha$ spanned
by the components whose coefficients are equal, we have envariance
over the entire subspace. For the rest of the space, it is only
relative phases of the components with unequal coefficients that can
be envariantly transformed.

\section{Allowed States}\label{AllowedStates}

The emergence of envariance is a reflection of the property of
entangled quantum states, where complete knowledge of the entire system
(i.e., the state being pure) means incomplete knowledge of the
subsystems. This can be understood in several ways:

1. The reduced density matrices, tracing out $\alpha$ or $\beta$ ($\mathrm{tr}_\alpha [\rho]$ or $\mathrm{tr}_\beta [\rho]$ for some pure $\rho$), have non-zero von Neumann entropy, leaving a mixed state partially
equivalent to a classical lack of knowledge about the subsystem. However, the number of invariant degrees of freedom is only indirectly related to the amount of entanglement as mentioned in \cite{Paris:2005ul}. We shall consider this point in more detail later on. 

2. Consider a Bell state,
\begin{equation}
\ket{\nu} = \tfrac{1}{\sqrt{2}}\left(\ket{\uparrow}_1\ket{\downarrow}_2 + \ket{\downarrow}_1\ket{\uparrow}_2\right),
\end{equation}
which is maximally entangled and as such has an SU(2) subgroup of
envariant transformations which we can parametrize by the Pauli
matrices,
\begin{equation}
e^{i\vec{\theta} \cdot \vec{\sigma}_1}\ket{\nu} = e^{i\vec{\theta} \cdot \vec{\sigma}_2}\ket{\nu},
\end{equation}
where $\sigma^i_j$ is the $i^{\textrm{th}}$ Pauli matrix for the
$j^{\textrm{th}}$ particle. Thus, rotating the spin of particle one is
the same as rotating the spin of particle two instead. This implies
only the {\it relative} orientations of the rays within the subsystem
Hilbert spaces are known. Zurek has cited a similar idea as his
motivation for using envariance \cite{Wheeler:1984dy} and calls it the
`relativity of quantum observables'. The situation can be said to have
a kind of Machianity \cite{Barbour:1994ri}, analogous to the situation where the
universe consists of point particles and only relative distances
between them are known, not global displacement or orientation. The state only contains information about the correlations between the
particles. 

3. For maximally entangled subsystems, the probabilities for fine grained (non-degenerate) measurement outcomes of a subsystem whose reduced density matrix has its maximum von Neumann entropy become equal {\it in any basis}. This can be seen with the use of envariance, which is equivalent to a basis ambiguity of the subsystems. For instance, with the Bell state the probabilities for a
particular particle to be spin up or down in the $z$-direction are the
same while the probabilities for the particle to be spin left or right
in the $x$ or $y$ direction are also the same. This is in contrast to
an unentangled spin-$1/2$ particle where there always exists a
direction where the spin is definitely known.

This last example is a special case of a phenomenon that does not
apply to classical physics. The uncertainty principle is usually
applied to pure states but the situation changes for mixed states,
such that the bounds on the uncertainties for incompatible observables
become more strict. To show this, we utilize the concavity of the
expression $-x\textrm{ln}x$ (with $x \in \mathbb{R}^+$) \cite{Wehrl:1978}
such that for a density matrix $\rho$ and a fine grained basis
$\left\{\ket{i}\right\}$, $\bra{i}j\rangle =
\delta_{ij}$, $\sum_i \ket{i}\bra{i} = \mathbb{I}$, the von Neumann
entropy $S(\rho)$ has the property:
\begin{align}
S(\rho) :&= -\mathrm{tr}[\rho\textrm{ln}\rho] \nonumber \\
&= \sum_i \bra{i}-\rho\textrm{ln}\rho\ket{i} \nonumber \\
&\leq -\sum_i \bra{i}\rho\ket{i}\textrm{ln}\bra{i}\rho\ket{i}.
\end{align}

Choosing $\left\{\ket{i}\left|i\right.\right\}$ to be the eigenstates
of a fine grained observable, then $\bra{i}\rho\ket{i}$ is the
probability to measure outcome ``$i$'' such that the Shannon entropy
of said observable (call it $O$) is given by
\begin{equation}
H_O(\rho) = -\sum_i \bra{i}\rho\ket{i}\textrm{ln}\bra{i}\rho\ket{i}.
\end{equation}
Thus,
\begin{equation}
S(\rho) \leq H_O(\rho).
\end{equation}
For an alternative proof see \cite{Schumacher:1996}.

This applies to all fine grained observables of the system described
by $\rho$. In the case of the Bell state, the reduced density matrix
obtained by tracing out one of the particles has maximum von Neumann
entropy such that all non-trivial observables of the subsystem also
have maximum Shannon entropy. Thus if a subsystem contains quantum
correlations with another, the information we have about the subsystem
is more constrained than in the case of classical physics where
Shannon entropies of `incompatible' observables are allowed to be
independent.

\section{Observable Dynamics of Completely Mixed States} \label{dynamics}

Intuitively, when one lacks knowledge of a system, one expects our
ability to distinguish the dynamics of the system to be lessened. We
have seen that in the case of mixed quantum states, our knowledge of
the system is less than allowed classically.  

We begin quantifying the distinguishability of dynamics of mixed
states by extending envariance of completely mixed states to
non-unitary operations. In this regard we choose to describe a quantum
process in an operator-sum representation which maps density matrices
to density matrices. A general physical operation on $\alpha$ can be
described by a set \cite{Nielsen:2004} of operation elements
$E_{k\alpha}\in \hil_\alpha\otimes\bar\hil_\alpha$ where
$\bar\hil_\alpha$ is the dual to $\hil$. The operation is then given
by
\begin{equation}
\mathcal{E}_\alpha(\rho) = \sum_{k=1}^K E_{k\alpha}\rho E^{\dagger}_{k\alpha}.
\end{equation}

We shall be concerned with operations $\mathcal{E}_\alpha$ whose
elements are trace preserving
$(\sum_k E_{k\alpha}^\dagger E_{k\alpha}=\mathbb{I}_\alpha)$ and also unital:
\begin{equation}
\sum_{k=1}^KE_{k\alpha}E^\dagger_{k\alpha}=\mathbb{I}_\alpha. \label{Condition}
\end{equation}
 Let $\rho_\alpha$ be a completely mixed state of a system $\alpha$,
 purified by system $\beta$,
\begin{equation}
\rho_\alpha := \mathrm{tr}_\beta[|\psi\rangle\langle\psi|] = \frac{1}{N}\mathbb{I}_\alpha.
\end{equation}
All unital operations leave $\rho_\alpha$ invariant (see Appendix A), i.e., 
\begin{equation}
\mathcal{E}_\alpha(\rho_\alpha) = \rho_\alpha.
\end{equation}

Some operations satisfying these conditions include:
\begin{enumerate}
\item Unitary $\mathcal{E}_U(\rho) = U\rho U^\dagger$.
\item Perfect premeasurements $\mathcal{E}_P(\rho) = \sum_{k=1}^{K}P_k \rho P_k$ where $P_k$ are projectors of a complete basis.
\item Combinations of unitary and perfect premeasurement operations, e.g., $\mathcal{E}_{UP}(\rho) = \mathcal{E}_U\circ\mathcal{E}_P(\rho)$ 
\end{enumerate}
Interestingly, generalized measurements \cite{Nielsen:2004} where the outcome is unknown do not necessarily satisfy these conditions, {\it e.g.}, for measurement operators of a two level system $M_1 = \ket{0}\bra{0}$ and $M_2 = \ket{0}\bra{1}$, the left hand side of Eq.~(\ref{Condition}) with $E_{i\alpha} = M_i$ does not equal unity; $M_1M^\dagger_1 + M_2M^\dagger_2 \neq \mathbb{I}$.

\section{Upper Bound for General Mixed States} \label{Upper}

Let us now consider a general mixed state of $\alpha$
\begin{equation}
\tilde{\rho}_{\alpha} = \sum_{j=1}^{n} |c_j|^2 |\alpha_j\rangle\langle\alpha_j|.
\end{equation}
A purification of $\alpha$ by $\beta$ is given by the Schmidt
decomposition (\ref{fourth}), where now the $c_i$ and $c_j$ may be
equal for $i \neq j$.  The group of envariant operations on
$|\Omega\rangle$ is in general greatly reduced compared to the
maximally entangled state.  Thus the set of all operations that can be
shown to leave $\tilde{\rho}_{\alpha}$ invariant by the use of
envariance is also reduced.  This limits the previous proof of the
unobservability of the dynamics of $\alpha$ for cases where the state
is not completely mixed.

Our proposal is that even with a large reduction in the set of
symmetries, the original set may apply in a partial sense.  The
motivation is that a large reduction in the symmetry can occur with
only a very small reduction in the von Neumann entropy of $\alpha$
\cite{Paris:2005ul}. Mixed states with less than maximum von Neumann
entropy may still have some form of limitations on their observable
dynamics for the full set of trace preserving operations satisfying
Eq.~(\ref{Condition}). This turns out to be the case.

To see this, we initially rewrite $\Omega$. Let us extend the
sum\footnote{However, the choice of the size of the extension may be
  chosen to be smaller depending on whether $\mathcal{E}$ leaves
  certain subspaces of $\hil_\alpha$ invariant.} over $j$ from one to
$N$ and define $c_j=0$ for $n + 1 \leq j \leq N$. For $M < N$, we
enlarge $\hil_\beta$ until the dimensionalities are equal. We then
decompose $\Omega$ into two parts, one that is maximally symmetric
over $\hil_\alpha$ and the rest of the state;
\begin{equation}
c_j = \frac{1}{\sqrt{N}} + d_j
\end{equation}
where $d_j := c_j - 1/\sqrt{N}$, such that
\begin{equation}
\ket{\Omega} = \frac{1}{\sqrt{N}} \sum^N_{j=1} e^{i\phi_j}\ket{\alpha_j}\ket{\beta_j} + \sum^N_{j=1}d_je^{i\phi_j}\ket{\alpha_j}\ket{\beta_j}.
\end{equation}
Define
\begin{align}
\ket{\Omega_1} &= \frac{1}{\sqrt{N}}\sum^N_{j=1}e^{i\phi_j}\ket{\alpha_j}\ket{\beta_j}, \\
Q\ket{\Omega_2} &= \sum^N_{j=1}d_j e^{i\phi_j}\ket{\alpha_j}\ket{\beta_j},
\end{align}
where the constant $Q=\sqrt{\sum_j d^2_j}$ is chosen such that
$\Omega_2$ is normalized to $1$. 

Our measure of the purity of $\alpha$ is given by $Q$. It is not equal to the usual measure of purity, which is $\mathrm{tr}[\tilde{\rho}^2_\alpha]$. One can
consider $Q^2$ as the $\chi^2$ value between the distribution of
amplitudes $\left\{c_j\right\}$ and the constant distribution
$1/\sqrt{N}$. It follows from $\sum_jc_j^2=1$ and $0\leq c_j^2\leq
c_j$ that $Q$ is bounded
\begin{equation}
0 \leq Q \leq \sqrt{2-2/\sqrt{N}.}
\end{equation}
The maximal value occurs for pure states of $\alpha$, while $Q=0$
corresponds to completely mixed states (all $c_j$ equal).

We now utilize a measure of the distinguishability of quantum states,
the trace distance, defined as
\begin{equation}
D(\rho,\sigma):=\tfrac{1}{2}\mathrm{tr}\vert\rho-\sigma\vert,
\end{equation}
where $\rho$ and $\sigma$ are density matrices and $\vert
X\vert:=\sqrt{X^\dagger X}$ is the positive square root of $X^\dagger
X$ (defined by taking a spectral decomposition $X^\dagger X =
\sum_ie_i\ket{x_i}\bra{x_i}$ and taking the positive square roots of
the eigenvalues $\sqrt{X^\dagger X} = \sum_i\sqrt{e_i}\ket{x_i}\bra{x_i}$). 

It can be shown that \cite{Nielsen:2004}
\begin{equation}
D(\rho,\sigma)=\operatorname*{\it max}_P \mathrm{tr}[P(\rho-\sigma)],
\end{equation}
where $P$ is a projector and the maximization is taken over all
possible projectors. This gives a clear physical interpretation of the
trace distance. If an experimentalist wanted to distinguish whether
they had the state $\rho$ or $\sigma$, the trace distance gives the
maximum possible difference in probabilities for a projective
measurement outcome for the two states. For
instance, if for two states $D=1$, it is in principle possible to do a
projective measurement where the probability of getting a confirmatory
result for one state is one while the other is zero and hence only one
measurement is ever needed to distinguish the states.

We are now in a position to derive an upper bound on $D_\alpha := D\big(\mathcal{E}_\alpha(\tilde{\rho}_\alpha),\tilde{\rho}_\alpha\big)$.

Let the state $\rho_{_\Omega} = \ket{\Omega}\bra{\Omega}$ be acted on by $\mathcal{E}_\alpha$ as defined in Sec. \ref{dynamics}.  In the case where $Q =
  0$, $\ket{\Omega} = \ket{\Omega_1}$ and
\begin{equation}
\mathcal{E}_\alpha\left(\rho_{_\Omega}|_{Q=0}\right) = \mathcal{E}_\beta\left(\rho_{_\Omega}|_{Q=0}\right),
\end{equation}
where\footnote{For cases where $\hil_\beta$ has to be enlarged and $\beta$ is
considered a real subsystem, $\mathcal{E}_\beta$ may not strictly be
physical.} $\mathcal{E}_\beta(\rho) = \sum_{k=1}^{K}E_{k\beta}\rho E^{\dagger}_{k\beta}$ (cf.\ Appendix A).  As these two states are equal, they are indistinguishable. For the
general case where $Q$ may not be zero, a measure for the
distinguishability of the two states can be given by the trace
distance
\begin{equation}
D_{\alpha\beta} := D\Big(\mathcal{E}_\alpha(\rho_{_\Omega}),\mathcal{E}_\beta(\rho_{_\Omega})\Big).
\end{equation}
In Appendix B, we show that $D_{\alpha\beta}$ satisfies the following bound:
\begin{equation}
D_{\alpha\beta} \leq 2\sqrt{1-\left| 1-Q^{2}+\tfrac{1}{4}Q^{4}\right|}.
\label{upperbound}
\end{equation}
This is related to $D_\alpha$ in the following way.  If $\beta$ is an ancilla subsystem used to purify $\alpha$ or the
experimentalist does not have access to subsystem $\beta$, then we can
ask about our ability to tell whether $\mathcal{E}_\alpha$ has happened
at all. This can be quantified by
\begin{align}
D\Big(\mathrm{tr}_\beta[\mathcal{E}_\alpha(\rho_{_\Omega})],\mathrm{tr}_\beta[\mathcal{E}_\beta(\rho_{_\Omega})]\Big) &= D\Big(\mathcal{E}_\alpha(\mathrm{tr}_\beta[\rho_{_\Omega}]),\mathrm{tr}_\beta[\rho_{_{_\Omega}}]\Big) \nonumber \\
&= D_\alpha.
\end{align}
The partial trace over $\beta$ is trace preserving, so $D_\alpha$ is bounded by $D_{\alpha\beta}$:
\begin{equation}
D_\alpha\leq D_{\alpha\beta}\leq2\sqrt{1-\left|1-Q^2+\tfrac{1}{4}Q^4\right|}.
\label{thirty-three}
\end{equation}
For values of $Q< \sqrt{2-\sqrt{3}} \approx 0.5$, the right hand side
of \eqref{thirty-three} becomes less than one and hence bounds
$D_\alpha$.  For $Q=0$, (\ref{thirty-three}) gives $D_\alpha=0$ which
is the same result achieved in Sec. \ref{dynamics}.  The upper bound
on $D_\alpha$ given by (\ref{thirty-three}) is our central result.

The non-trivial nature of this bound can be seen by considering cases where $D_\alpha$
is not bounded because $Q$ is larger than $\sqrt{2-\sqrt{3}}$:

\begin{enumerate}
\item Consider two bases for $\alpha$, $\{|\alpha_k\rangle\}$ and $\{|\tilde{\alpha}_k\rangle\}$ such that $\langle\alpha_m|\tilde{\alpha}_m\rangle = 0$ for some $m$.  Take a pure state $\sigma = |\alpha_m\rangle\langle\alpha_m|$.  The purity as given by $Q$ is then,
\begin{equation}
Q = \sqrt{2 - 2/\sqrt{N}} > \sqrt{2 - \sqrt{3}}
\end{equation}
for $N \geq 2$.  One can see that if $\alpha$ experiences a unitary transformation
\begin{equation}
U_\alpha = \sum_k |\tilde{\alpha}_k\rangle\langle\alpha_k|,
\end{equation}
then the states have zero overlap:
\begin{equation}
\langle \alpha_m|U_\alpha\sigma U^\dagger_\alpha|\alpha_m\rangle \quad \mbox{or} \quad D(U_\alpha\sigma U^\dagger_\alpha,\sigma) = 1.
\end{equation}
Thus the two states are in principle easily distinguishable.

\item Suppose the system $\alpha$, still given by the pure state
  $\sigma$, experiences a perfect premeasurement such that
\begin{equation}
\sigma \rightarrow \sum^N_{i=1}P_i\sigma P_i,
\end{equation}
where $P_i = \ket{A_i}\bra{A_i}$ and $\left\{\ket{A_i}\right\}$ forms a complete orthonormal basis for
$\hil_\alpha$ such that $\ket{\alpha_m} =
\sum^N_{i=n}(1/\sqrt{N})\ket{A_i}$. It is convenient to use the
definition of fidelity \cite{Nielsen:2004} for density
matrices $\rho$ and $\tau$
\begin{equation}
F(\rho,\tau):=\mathrm{tr}\big[\sqrt{\rho^{1/2}\tau\rho^{1/2}}\big],
\end{equation}
to obtain
\begin{align}
\big(F(\sigma,\sum_iP_i\sigma P_i)\big)^2 &= \bra{\alpha_m}\sum_iP_i\sigma P_i\ket{\alpha_m} \nonumber \\
&= \frac{1}{N}.
\end{align}
In this case, the fidelity bounds the trace distance
\begin{align}
& 1 - \big(F(\sigma,\sum_iP_i\sigma P_i)\big)^2 \nonumber \\
&= 1 - \frac{1}{N} \leq D(\sigma,\sum_iP_i\sigma P_i).
\end{align}
Thus, the observability of the process $D(\sigma,\sum_i P_i\sigma P_i)$ tends to 1 as $N$ tends to $\infty$.
\end{enumerate}

Finally, we note that the bound may be extended to mixed states of a
composite $\alpha$ and $\beta$ system. Let
\begin{equation}
\bar{\rho} = \sum_m r_m\rho_m,
\end{equation}
where $\sum_mr_m = 1$ and $\rho_m$ is a pure density matrix of the
composite system $\forall m$. Define $Q_m$ as the $Q$ measure of the
purity of the $\mathrm{tr}_\beta(\rho_m)$ state and define
$\mathcal{E}_\alpha$ and $\mathcal{E}_\beta$ in the usual way. Then, using
the convexity of the trace distance,
\begin{align}
D\Big(\mathcal{E}_\alpha(\bar{\rho}),\mathcal{E}_\beta(\bar{\rho})\Big) &\leq \sum_m r_m D\Big(\mathcal{E}_\alpha(\rho_m), \mathcal{E}_\beta(\rho_m)\Big) \nonumber \\
&\leq 2\sum_m r_m \sqrt{1 - \left|1 - Q^2_m + \tfrac{1}{4}Q^4_m\right|}.
\end{align}

Thus, the distinguishability of the dynamics of $\alpha$ in a mixed
composite state is bounded by the average of the bounds of the pure
states $\rho_m$.

\section{Remarks}

We have shown that given a trace preserving unital operation, the trace distance between the
transformed state and its original is bounded by (\ref{thirty-three})
given that the purity is enough ($Q\lesssim 0.5$).
For maximally mixed states where $Q=0$, the bound implies the
operation must be unobservable.

The bound (\ref{thirty-three}) is motivated in Sec.~\ref{AllowedStates} and Sec.~\ref{dynamics} on the intuition that lack of knowledge of a state leads to lack of an ability to distinguish the dynamics. We note that trace preserving, unital operations cannot decrease the von Neumann entropy \cite{Garcia:2006}. This leads us to ask whether the class of trace preserving, unital operations is the largest such class where (\ref{thirty-three}) or a stronger bound holds that depends only on the purity of the input state.

\begin{acknowledgments}
I thank R.J. Crewther and Lewis C. Tunstall for a critical reading of the manuscript.  I acknowledge technical support from S. Underwood, Dale S. Roberts, and A. Casey.  This work is supported by the Australian Research Council.
\end{acknowledgments}

\setcounter{secnumdepth}{-1}
\appendix
\section{Appendix A: Symmetries of Completely Mixed States}
\label{app:A}
The proof of the invariance of $\rho_\alpha$ under unital operations
is trivial. Here we provide an alternative proof which give the tools
needed for Sec.~\ref{Upper}. The first step is to extend the symmetry
of the second version of (\ref{envdef}) for pure state (\ref{third}),
to $\mu_\alpha |\psi\rangle = \mu_\beta |\psi\rangle$ where
$\mu_\alpha := \bar{\mu}_\alpha\otimes\mathbb{I}_\beta$ is a general
linear operation on pure states.  The operations $\bar{\mu}_\alpha$
could for instance be a projector onto a subspace of
$\mathcal{H}_\alpha$.  Also suppose $\ket{\psi}$ is maximally
entangled with respect to subsystems $\alpha$ and $\beta$.  Since
$\mu_\alpha$ acts identically on subsystem $\beta$, it follows that
\begin{equation}
\mu_\alpha \ket{\psi} = \frac{1}{\sqrt{N}}\sum^N_{j=1}e^{i\phi_j}\left(\bar{\mu}_\alpha\ket{\alpha_j}\right)\ket{\beta_j}.
\end{equation}
Define $\bra{\alpha_i}\bar{\mu}_\alpha\ket{\alpha_j} := \mu_{ij}$ such that
\begin{align}
\mu_\alpha\ket{\psi} &= \frac{1}{\sqrt{N}}\sum^N_{j=1}e^{i\phi_j}\Big(\sum^N_{i=1}\mu_{ij}\ket{\alpha_i}\Big)\ket{\beta_j} \nonumber \\
&= \frac{1}{\sqrt{N}}\sum^N_{i=1}\ket{\alpha_i}\Big(\sum^N_{j=1}e^{i\phi_j}\mu_{ij}\ket{\beta_j}\Big) \nonumber \\
&= \frac{1}{\sqrt{N}}\sum^N_{i=1}\ket{\alpha_i}\ket{\tilde{\beta}_i},
\end{align}
where $\ket{\tilde{\beta}_i} := \sum^N_{j=1}e^{i\phi_j}\mu_{ij}\ket{\beta_j}$. For non-unitary $\mu_\alpha$, $\{\ket{\tilde{\beta}_j}\}$ need not be orthonormal.\footnote{Many $|\tilde{\beta}_j\rangle$ may even be the null state.}

Define
\begin{align}
\mu_\beta:=\mathbb{I}_\alpha&\otimes\sum^N_{k=1}e^{i\phi_k}\ket{\tilde\beta_k}\bra{\beta_k}:=\mathbb{I}_\alpha\otimes\bar\mu_\beta, \\
\therefore\mu_\alpha\ket{\psi}&=\mu_\beta\ket{\psi}. \label{16}
\end{align}
This symmetry is not equivalent to envariance as $\mu_\beta$ may not be invertible and hence in general there does {\it not} exist a $\mu_\beta^{-1}$ such that $\mu_\alpha\mu_\beta^{-1}|\psi\rangle = |\psi\rangle$.  With Eq. (\ref{16}), we can consider symmetry properties of completely mixed states.  Let the completely mixed state to be a state of system $\alpha$;
\begin{equation}
\rho_\alpha := \mathrm{tr}_\beta [\ket{\psi}\bra{\psi}] = \frac{1}{N}\mathbb{I}_\alpha.
\end{equation}
Suppose we have a quantum operation $\mathcal{E}_\alpha(\rho)=\rho'$ that is given in an operator-sum representation;
\begin{equation}
\mathcal{E}_\alpha(\rho) = \sum^K_{k=1}E_{k\alpha}\rho E^{\dagger}_{k\alpha},
\end{equation}
where $E_{k\alpha}$ are linear maps $E_{k\alpha} : \hil_\alpha
\rightarrow \hil_\alpha$. The effect of $\mathcal{E}_\alpha$ upon $\rho_\alpha$ is
then
\begin{align}
\mathcal{E}_\alpha(\rho_\alpha) &= \sum^K_{k=1} E_{k\alpha}\mathrm{tr}_\beta\left[\ket{\psi}\bra{\psi}\right]E^{\dagger}_{k\alpha} \nonumber \\
&= \mathrm{tr}_\beta\Big[\sum^K_{k=1}E_{k\alpha}\ket{\psi}\bra{\psi}E^{\dagger}_{k\alpha}\Big].
\end{align}
Since $E_{k\alpha}$ is of the form $\bar{\mu}_\alpha$, there exists an
$E_{k\beta}$ such that $E_{k\alpha}\ket{\psi} = E_{k\beta}\ket{\psi}$
and similarly $\bra{\psi}E^\dagger_{k\alpha} = \bra{\psi}E^\dagger_{k\beta}$. Then
\begin{align}
\mathcal{E}_\alpha(\rho_\alpha) &= \mathrm{tr}_\beta\Big[\sum^K_{k=1}E_{k\beta}\ket{\psi}\bra{\psi}E^\dagger_{k\beta}\Big] \nonumber \\
&= \mathrm{tr}_\beta\Big[\sum^K_{k=1}E^\dagger_{k\beta}E_{k\beta}\ket{\psi}\bra{\psi}\Big].
\end{align}
If $\sum^K_{k=1}E^\dagger_{k\beta}E_{k\beta} = \mathbb{I}_\beta$, then
\begin{equation}
\mathcal{E}_\alpha(\rho_\alpha) = \rho_\alpha.
\end{equation}
The completely mixed state is then invariant under $\mathcal{E}_\alpha$
if $\mathcal{E}_\alpha$ satisfies the condition $\sum^K_{k=1}E^{\dagger}_{k\beta}E_{k\beta} = \mathbb{I}_\beta$, which is equivalent to the unital condition
\begin{equation}
\sum^K_{k=1}E_{k\alpha}E^{\dagger}_{k\alpha} = \mathbb{I}_\alpha.
\end{equation}

\section{Appendix B: Proof of upper bound of trace distance}
\label{app:B}
To prove Eq.~(\ref{upperbound}), we shall need a few identities. Using
the definition of $d_i$ and taking $\Omega$ to be normalized, we find;
\begin{align}
\sum^N_{i=1}c^2_i &=1=\sum^N_{i=1}\left(\frac{1}{\sqrt{N}}+d_i\right)^2 \nonumber \\
&=1+\sum^N_{i=1}d_i^2+\frac{2}{\sqrt{N}}\sum_{i=1}^Nd_i \nonumber \\
\therefore \sum_{i=1}^Nd_i^2 &=-\frac{2}{\sqrt{N}}\sum^N_{i=1}d_i
\label{coefficients}
\end{align}
We also compute the overlap
\begin{equation}
\bra{\Omega_1}\Omega_2\rangle=\frac{1}{Q\sqrt{N}}\sum_{i=1}^Nd_i=-\frac{1}{2}Q,
\end{equation}
using Eq.~(\ref{coefficients}). Defining
\begin{align}
\rho_{11}&:=\ket{\Omega_1}\bra{\Omega_1} \nonumber \\
\rho_{12}&:=\ket{\Omega_1}\bra{\Omega_2} \nonumber \\
\rho_{21}&:=\ket{\Omega_2}\bra{\Omega_1} \nonumber \\
\rho_{22}&:=\ket{\Omega_2}\bra{\Omega_2},
\end{align}
we have
\begin{align}
\rho &= \ket{\Omega}\bra{\Omega} \nonumber \\
&= \rho_{11}+Q(\rho_{12}+\rho_{21})+Q^2\rho_{22}.
\end{align}
The $\rho$s have some useful relationships, namely,
\begin{align}
\rho_{11}^2 &= \rho_{11}, \rho_{22}^2 = \rho_{22}, \nonumber \\
\rho_{12}\rho_{21} &= \rho_{11}, \rho_{12}\rho_{12} = -\tfrac{Q}{2}\rho_{12}, \nonumber \\
\rho_{21}\rho_{12} &= \rho_{22}, \rho_{21}\rho_{21} = -\tfrac{Q}{2}\rho_{21}.
\end{align}
The trace distance has particular properties which shall also be
used. Thus,
\begin{align}
D_{\alpha\beta}:&=D\Big(\mathcal{E}_\alpha(\rho),\mathcal{E}_\beta(\rho)\Big) \nonumber \\
&\leq D\Big(\mathcal{E}_\alpha(\rho),\mathcal{E}_\alpha(\rho_{11})\Big)+D\Big(\mathcal{E}_\alpha(\rho_{11}),\mathcal{E}_\beta(\rho)\Big) \nonumber \\
&=D\Big(\mathcal{E}_\alpha(\rho),\mathcal{E}_\alpha(\rho_{11})\Big)+D\Big(\mathcal{E}_\beta(\rho_{11}),\mathcal{E}_\beta(\rho)\Big) \nonumber \\
&\leq D(\rho,\rho_{11})+D(\rho_{11},\rho) \nonumber \\
&=2D(\rho_{11},\rho).
\end{align}
Because $\rho$ and $\rho_{11}$ are pure,
\begin{equation}
D(\rho_{11},\rho)=\sqrt{1-\big(F(\rho_{11},\rho)\big)^2},
\end{equation}
where $F$ is the fidelity;
\begin{align}
F(\rho_{11},\rho) &= \mathrm{tr}\Big[\big(\rho_{11}^{1/2}\rho\rho_{11}^{1/2}\big)^{1/2}\Big] \nonumber \\
&=\mathrm{tr}\Big[\big(\rho_{11}\rho\rho_{11}\big)^{1/2}\Big] \nonumber \\
&=\mathrm{tr}\Big[\big(\rho_{11}[\rho_{11}+Q(\rho_{12}+\rho_{21})+Q^2\rho_{22}]\rho_{11}\big)^{1/2}\Big] \nonumber \\
&=\mathrm{tr}\Big[\big(\rho_{11}-\tfrac{1}{2}Q^2\rho_{11}-\tfrac{1}{2}Q^2\rho_{11}+\tfrac{1}{4}Q^4\rho_{11}\big)^{1/2}\Big] \nonumber \\
&=\sqrt{1-Q^2+\tfrac{1}{4}Q^4}.
\end{align}
Thus,
\begin{equation}
D_{\alpha\beta}\leq 2\sqrt{1-\left|1-Q^2+\tfrac{1}{4}Q^4\right|}.
\end{equation}


\end{document}